\documentclass[12pt,a4paper]{article}
\pdfoutput=1
\usepackage{jheppub}
\usepackage{amsmath,amssymb,scalefnt}
\usepackage{xfrac}
\usepackage{graphicx}
\usepackage{caption}
\usepackage{subcaption}

\PassOptionsToPackage{jheppub}{hyperref}

\makeatletter
\everymath{\if@display\else\thickmuskip=4mu plus 4mu\fi}
\makeatother

\newcommand{\beq}{\begin{equation}}
\newcommand{\eeq}{\end{equation}}
\newcommand{\bea}{\begin{eqnarray}}
\newcommand{\eea}{\end{eqnarray}}

\def\ocal{{\cal O}}

\def\scal{{\cal S}}

\def\pa{\partial}
\newcommand{\vev}[1]{\left\langle#1\right\rangle}
\newcommand{\eqn}[1]{Eq.~{\hspace{-2pt}}(\ref{#1})}

\newcommand{\secn}[1]{Sec.~\hspace{-2pt}\ref{#1}}

\newcommand{\reference}[1]{Ref.~\hspace{-2pt}\cite{#1}}

\newcommand{\norm}[1]{\lVert#1\rVert}

\def\ab{{\alpha\beta}}
\def\ge{{\gamma\epsilon}}
\def\mn{{\mu\nu}}
\def\kl{{\kappa\lambda}}

\preprint{\hskip0.15in NSF-KITP-16-085,\ MCTP-16-14,\ LTH 1088}

\title{Zero modes in de~Sitter background}

\author[a,1]{Martin B Einhorn}\note{Also, \it{Michigan Center for 
Theoretical Physics, Ann Arbor, MI 48109.}} 
\author[a,b]{and D R Timothy Jones}
\affiliation[a]{Kavli Institute for Theoretical Physics,\\ University of California,
Santa Barbara, CA 93106-4030, USA}
\affiliation[b]{Dept. of Mathematical Sciences,\\ University of Liverpool, 
Liverpool L69 3BX, UK}

\emailAdd{meinhorn@umich.edu}
\emailAdd{drtj@liv.ac.uk}

\abstract{
There are five well-known zero modes among the fluctuations of the 
metric of de~Sitter (dS) spacetime. For Euclidean signature, they can 
be associated with certain spherical harmonics on the $S^4$ sphere, 
viz., the vector representation $\bf5$ of the global $SO(5)$ isometry. 
They appear, for example, in the perturbative calculation of the on-shell 
effective action of dS space, as well as in models containing matter 
fields. These modes are shown to be associated with collective modes 
of $S^4$ corresponding to certain coherent fluctuations.  When dS 
space is embedded in flat five dimensions $E^5,$ they may be seen as 
a legacy of translation of the center of the $S^4$ sphere.  
Rigid translations of the $S^4$-sphere on $E^5$ leave the classical 
action invariant  but are unobservable displacements from the point of 
view of gravitational dynamics on $S^4.$ Thus, unlike similar moduli, 
the center of the sphere is not promoted to a dynamical degree of 
freedom.  As a result, these zero modes do not signify the possibility of 
physically realizable fluctuations or flat directions for the metric of dS 
space. They are not associated with Killing vectors on $S^4$ but can be 
identified with certain non-isometric, conformal Killing forms that locally 
correspond to a rescaling of the volume element $dV_4.$ 

We frame much of our discussion in the context of 
renormalizable gravity, but, to the extent that they only depend upon
the global symmetry of the background, the conclusions should apply 
equally to the corresponding zero modes found in Einstein gravity.  
Although their existence 
has only been demonstrated at one-loop, we expect that these zero 
modes will be present to all orders in perturbation theory. They will 
occur for Lorentzian signature as well, so long as the hyperboloid 
$H^4$ is locally stable, but there remain certain infrared issues that 
need to be clarified. We conjecture that they will appear in any 
gravitational theory having dS background as a locally stable solution of 
the effective action, regardless of whether additional matter is included.}

\arxivnumber{1606.02268v4}

\begin{document}
\maketitle

\section{Introduction}
\label{sec:intro}

There are five well-known zero modes in the conformal fluctuations of
the metric of de~Sitter (dS) space. For Euclidean signature, they are
associated with the spherical harmonics on the sphere $S^4$ 
corresponding to the vector representation $\bf5$ of $SO(5).$  
These five zero modes are ubiquitous, appearing in renormalizable
gravity, both with and without additional matter, as well as in
loop corrections to the usual Einstein-Hilbert (E-H) theory, treated as
an effective field theory. $S^4$ may be embedded in flat
five-dimensional spacetime $E^5,$ whose isometries are the Poincar\'e
group, $SO(5){\rtimes}P^5.$  We shall show that the invariance of the 
embedding under translations in five-dimensions $(P^5)$ is reflected by 
certain collective modes or moduli that leave the gravitational action in four-
dimensions invariant. In the transverse-traceless gauge, these can be 
associated with certain conformal fluctuations of the metric on $S^4.$  

For Lorentzian signature, a similar analysis is expected to apply to the 
hyperboloid $H^4$ with isometry $SO(4,1),$ although there are 
subtleties that have not been resolved stemming from the long-range 
behavior of the fluctuations. We do not believe this infrared issue 
represents an insuperable obstacle to analytic continuation from 
Euclidean to Lorentzian signature. 

These zero modes appear
to be a universal feature of models in dS space,  for reasons that will
be explained in this paper.   Our point of view regarding Euclidean
quantum gravity is more or less  the same as that expressed by  
Christensen \& Duff~\cite{Christensen:1979iy}, except that we extend  
that philosophy to renormalizable gravity. The E-H theory has well known
instabilities in  the conformal sector, and it has been 
suggested~\cite{Gibbons:1978ac} that the contour of integration in the 
Euclidean path integral (EPI) be changed for these unstable modes.  Even
if one adopts their prescription, these five
zero modes persist.  However, some may take the point of view that the
entire framework is suspect  as a result of those instabilities.  One advantage of renormalizable gravity is
that,  with a sensible choice of the sign of the coupling constants,
there is no need to  modify the definition of the EPI to achieve
convergence in the conformal sector.  Further, for a subset of this
range of couplings, there are no unstable modes for  fluctuations about
dS background at one-loop  order~\cite{Avramidi:1986mj,
Avramidi:2000bm}. Nevertheless,  there remain the five zero modes that
are the focus of this paper. A second advantage of renormalizable gravity is that 
it is asymptotically free in the  gravitational
couplings~\cite{Fradkin:1981hx, Fradkin:1981iu, Avramidi:1985ki,
Avramidi:2000bm, Avramidi:1986mj, Salvio:2014soa}.  In certain
circumstances, asymptotic freedom may be extended to all 
couplings\footnote{The work by Buchbinder and collaborators is reviewed
in detail in Chapter~9 of
\reference{Buchbinder:1992rb}.}~\cite{Buchbinder:1986cx,
Buchbinder:1989jd, Buchbinder:1989ma, Einhorn:2016mws}.  As a result,
perturbation theory can be trusted at sufficiently high scales.

As did the  authors of \reference{Christensen:1979iy}, we reject the
notion that spacetime  is asymptotically flat, since that is not a
solution of the field equations  in the presence of a nonzero
cosmological constant. Correspondingly, we  cannot assume the existence
of an S-matrix but instead emphasize  correlation functions through the
perturbative calculation of the  effective action $\Gamma[g_\ab].$ (For
the same reason, it is also  important not to discard the Gauss-Bonnet 
term~\cite{Christensen:1979iy, Einhorn:2014bka},  which is nonzero at
every point.)  

Some of our results overlap with a paper by 
Gibbons \& Perry\footnote{\reference{Christensen:1979iy} 
corrects some numerical errors in \reference{Gibbons:1978ji}, but our 
results do not depend upon such~details.}
\cite{Gibbons:1978ji}. In particular, in their Sec.~2, they cite 
a theorem~\cite{Yano:1959} that, assuming Euclidean signature, these 
non-isometric, conformal zero modes can only occur in $d=4$ for $S^4.$ 
(Surprisingly, these modes are passed over in their 
treatment of Euclidean dS space in their Sec.~4.)
The extension of this theorem to the 
pseudo-Riemannian case has been addressed subsequently.  Assuming 
Einstein's equations in vacuum, a much stronger assumption than 
assuming that  spacetime is an Einstein space, the theorem can be 
extended to Lorentzian signature~\cite{Garfinkle:1986vu}, implying the 
spacetime is either dS or anti-de~Sitter (AdS). This topic has also received 
further attention in the mathematical literature; for a recent review and 
discussion, see \reference{Kuhnel:2009}. Under various technical assumptions, 
much weaker than requiring Einstein's equations, such non-isometric, conformal 
zero modes for Lorentzian signature can only occur in a spacetime of constant 
curvature.  In general, the manifold need not be simply connected, so it cannot 
be inferred that the spacetime is dS or AdS without additional assumptions.

Folacci~\cite{Folacci:1992xc, Folacci:1996dv} has also discussed the
nature of these five zero modes, suggesting that they should be regarded
as gauge artifacts. In \reference{Christensen:1979iy}, it is shown that
one obtains agreement between the general result in an arbitrary
background and the dS result only if these zero modes are counted. Since
these modes are present on-shell, it seems doubtful that they are true
gauge artifacts, but their unphysical nature does resemble a gauge
symmetry. For Lorentzian signature, although unproven, our point of view
is also different from his, an issue to which we shall return in
\secn{sec:lorentz}.

In the next section, we review some aspects of the background field
method for calculating the effective action in perturbation theory.
Then, in \secn{sec:ds5}, we describe the embedding of Euclidean dS space
as a submanifold in flat five dimensions. In \secn{sec:lift}, we do the
reverse, explaining how one may lift metric fluctuations from four to
five dimensions. Finally, in \secn{sec:5zero}, we interpret the five
zero modes as remnants of a potential collective mode in five
dimensions. Some comments concerning the extension to Lorentzian
signature are contained in \secn{sec:lorentz}. Finally, further
discussion and conclusions follow in Sections~\ref{sec:discuss} and
\ref{sec:conclude}, respectively.

\section{Effective action}\label{sec:effact}

The calculation of the effective action\footnote{There are numerous effective 
actions that have been defined. In this paper, we shall only employ  the 
generating functional $\Gamma$ of 1PI Green's functions, the Legendre 
transform of $W[J].$  It is gauge-dependent although its value at an 
extremum is not.} of a quantum field theory (QFT) is one of the most useful 
ways to explore its properties. The perturbative calculation of the effective 
action in quantum gravity has a long 
history\footnote{See 
\reference{Buchbinder:1992rb, Avramidi:1986mj, Avramidi:2000bm} for 
extensive reviews.}.  In general, it is technically complicated by the large 
gauge symmetry (diffeomorphism invariance) and tensorial calculus, as well as 
by the conceptual issues associated with the fact that, in a sense, the 
spacetime itself is determined self-consistently by the calculation.

Our interest was stimulated in part by Avramidi's 
calculation~\cite{Avramidi:1986mj, Avramidi:2000bm}  of the effective
potential for the curvature of dS space in renormalizable gravity,  but
our result concerning these five zero modes depends only  upon the
symmetries of the background field.  Conceptually, it is somewhat 
simpler to begin with a renormalizable theory in which the EPI is
well-defined.  Like Avramidi, we can assume that the action has both
an  E-H term as well as a cosmological constant, assumed positive. 
(There are slight but important changes required to accommodate the 
classically scale invariant cases of interest in our other recent 
work~\cite{Einhorn:2014gfa, Einhorn:2015lzy, Einhorn:2016mws}.)  We
shall  deal here with the purely gravitational case, but it will be
self-evident that it can  be generalized to the inclusion of matter
fields in dS background.

 The calculation of the effective action
 \beq \label{eq:effaction}
 \Gamma_{eff}[g_\ab(x)]\equiv S_{cl}[g_\ab(x)]+
 {\Delta\Gamma}_{eff}[g_\ab(x)]
 \eeq
 in perturbation theory is somewhat simplified by the background field method, 
 which is most easily described in terms of the EPI. 
 To establish notation, the classical action will be expressed as 
\beq\label{eq:scl}
S_{cl}\! =\!\! \int\!\! d^4x\sqrt{g}\left[ 
\frac{C^2}{2a}+ \frac{R^2}{3b}+ c\, G-
\frac{M_P^2}{2}\left(R-2\Lambda\right)\right],
\eeq
where $C \equiv C^\gamma{}_{\epsilon\ab}$ is the Weyl tensor; $G,$ the Gauss-Bonnet term 
$G\equiv C^2-2W;$ $W\equiv R_\ab^2-R^2/3;$ $R_\ab,$ the Ricci tensor; and $R,$ the 
scalar curvature. $\!M_P$ is the reduced Planck mass or string scale, and 
$\Lambda$ is the cosmological constant, both assumed positive. The maximally 
symmetric solution of the classical field equations has the same form as in E-H 
gravity, viz., $R_\ab=\Lambda g_\ab,$ with a metric $g_\ab^{dS}(x)$ that 
describes dS space. (Its application to cosmology requires certain 
additional assumptions that will not be taken up here.)
 
We shall next summarize the use of the background field method to calculate 
the effective action perturbatively\footnote{In the present context, a brief 
overview can be found in an appendix to \reference{Einhorn:2014gfa}.}.  Our 
purpose for reviewing this is to clarify what the fields $h_\ab$ represent from 
the point of view of the EPI, which, because the theory is required to be 
invariant under arbitrary diffeomorphic transformations of the metric, is most 
clearly expressed in the language of differential geometry. (Readers familiar 
with the effective action and with the background field method may skip to the 
next section.) To proceed, one places this classical action in the EPI and 
attempts to integrate over all metrics $g_\ab(x)$ under certain consistency 
conditions.  Convergence of the EPI requires that the couplings $a,b>0$ (with 
the G-B coupling $c$ determined by $a,b$ up to an additive 
constant~\cite{Einhorn:2014bka}). To perform the EPI, the metric is split 
 \beq
g_\ab(x) \equiv g^B_\ab(x)+h_\ab(x),
\eeq
where the ``classical" background field $g^B_\ab(x)$ is generically an arbitrary 
function to be determined, and the quantum field $h_\ab(x)$ is to be integrated 
out. $h_\ab(x)$ will be referred to as the quantum fluctuations or simply 
fluctuations.  The Feynman rules for $h_\ab(x)$ are obtained in principle by 
expanding $S[g^B_\ab(x)+h_\ab(x)]$ in powers of $h_\ab(x)$ and dropping 
the linear term. The consistency condition mentioned above is that the 
one-point function for $h_\ab$ vanish to all orders, i.e., the classical field is in 
fact the background field, $\vev{g_\ab}=g^B_\ab.$  
Said otherwise, the effective action is the generating functional of 
one-propagator-irreducible\footnote{Given that the notion of ``particle" is frame 
dependent and the propagator refers to the quanta associated with fluctuations 
$h_\ab$ in a certain background field, this appellation seems more appropriate 
than the usual nomenclature ``one-particle-irreducible."} (1PI) Feynman diagrams 
in the presence of a classical background $g^B_\ab(x).$  Its extrema,
\beq\label{eq:onshell}
\frac{\delta \Gamma_{eff}[g^B_\ab]}{\delta g^B_\ab(x)}=0,
\eeq
replace the classical equations of motion (EOM).
When \eqn{eq:onshell} is satisfied, the background field is said to be 
``on-shell."  Stability of the solution is investigated by evaluating higher-order 
variations on-shell.  In principle, this should be carried out for an arbitrary 
background field but, in practice, it is often restricted by certain assumptions 
about the relevant global isometries.

This seemingly circular procedure for determining $g^B_\ab$ is, in principle,  
straightforward to implement in perturbation theory, although in practice,
it can seldom be carried out explicitly without further approximations. In 
lowest order, the background field is approximated by the solution of the 
(renormalized) classical field equations $g^B_\ab(x) {\to} g_\ab^{cl}(x)$, 
which may receive quantum corrections in higher order. Following the 
procedure described above, the propagator and vertices depend explicitly on 
the background field.  The one-loop result, which is as far as these calculations 
have been carried in renormalizable quantum gravity, 
is determined by the quadratic terms alone, which take the form of a 
functional integral over $\exp[-h_\ge(x)\ocal^{\ge\ab}h_\ab(x)],$ for some local 
differential operator 
$\ocal^{\ge\ab}(g_{\rho\sigma}^B(x),\nabla_\tau{}^B),$ where $\nabla^B$ 
represents the covariant derivative associated with the background metric.

Although the fluctuations $h_\ab(x)$ have no particular symmetry, in the
case  at hand, the operators $\ocal^{\ge\ab}(x)$ will be restricted by
the $SO(5)$  global symmetry presumed of the background.  In order to
carry out  the integration, it must be that the eigenvalues of 
$\ocal^{\ge\ab}(g_{\rho\sigma}^B(x),\nabla_\tau{}^B),$  are
non-negative, at least in a neighborhood of being on-shell. Even
if there are no truly unstable fluctuations, there may occur certain 
``flat directions" or ``zero modes", i.e., field configurations
$h^{(z)}_\ab(x)$ that  make no change in the value of the action. To
one-loop order, these can be  expressed by the partial differential
equation
\beq\label{eq:o2pde}
\ocal^{\ge\ab}\big(g_\kl^{cl}(x);\nabla^{cl}_\tau\big)h^{(z)}_\ab(x){=}0,
\eeq  
in which the background field is taken to be a solution to the classical EOM.
 
The occurrence of zero modes in QFT is seldom 
accidental and usually reflects some symmetry of the theory, either unbroken 
or spontaneously broken, or the identification or emergence of some 
collective coordinate. In the case of spontaneous broken gauge theories, 
massless Goldstone bosons persist unless absorbed by giving mass to some 
vector bosons. Sometimes, pseudo-Goldstone bosons occur because of 
some symmetry of the dimension-four scalar interactions that is not a symmetry 
of the full QFT~\cite{Weinberg:1972fn}.
In principle, one may try to calculate two-loop and higher
corrections to determine whether these zero modes persist, but, in these 
gravitational models, it is usually prohibitively complicated to carry out. 

Local gauge symmetries complicate the issue further because they
guarantee that certain transformations of a vector field or metric are
physically equivalent and have no effect on-shell. This implies that the
fluctuations may be subdivided into equivalence classes wherein each
element is a gauge-transform of another. Aside from lattice gauge theory, 
the only way found so far to deal with this redundancy is to select a 
single representative or a subset of representatives of each equivalence class 
by means of ``gauge-fixing" constraints that allow the propagator to be 
determined and to add so-called Faddeev-Popov ghost fields  to ensure that the 
final result is independent of the representative chosen\footnote{In fact, this 
technique has evolved into ``gauge-averaging" rather than gauge-fixing. We
have not seen this treated in the mathematical physics literature.}. As
if this were not confusing enough, in gravity, the very choice of a
coordinate system in which to express the background metric 
$g_\ab^B(x)$ already involves at least a partial choice of gauge.  Unfortunately,
for curved spacetime, no coordinate-independent method of calculation has been found.
To calculate to higher-order, after gauge-fixing, the quadratic form involving a 
modified operator $\widetilde{\ocal}^{\ge\ab}(\nabla^B;g_\kl^B)$ 
is inverted to define propagators for $h_\ab(x),$ and the terms cubic and 
higher-order in $h_\ab$ determine the ``interaction  vertices".  Although the 
initial form of the second-order terms does not require gauge-fixing or even a 
specific choice of coordinates, the actual evaluation of the functional 
determinants arising at one-loop and the calculations at two-loops and 
higher do. Consequently, in general, the result for $\Gamma_{eff}$ will depend 
on the choice of gauge; however, ``on-shell,"  when \eqn{eq:onshell} is satisfied, 
this gauge dependence must disappear.  As summarized above, to one-loop 
order, the first approximation to these on-shell conditions correspond simply to 
solutions of the classical EOM.

Even so, as in ordinary QFT in Minkowski space, the one-loop effective action 
cannot be evaluated analytically (or numerically) except in certain very special 
backgrounds. In non-gravitational models, in the case of spacetime 
independent background fields, the effective action reduces to an effective 
potential, whose generic form is known. The most nearly analogous case in 
gravity corresponds to a maximally-symmetric background metric, such as dS 
or AdS, together with constant matter fields, if present.  Unfortunately, the 
generic form of the one-loop potential is not known in this case. Nevertheless, 
the one-loop calculation in dS for pure higher-derivative gravity has been 
carried out in certain cases, and the beta-functions, which \emph{are} gauge
independent, have been determined in 
general~\cite{Fradkin:1981hx, Fradkin:1981iu, Avramidi:1985ki, 
Avramidi:1986mj, Salvio:2014soa}.   In particular, 
Avramidi~\cite{Avramidi:1986mj,Avramidi:2000bm} showed that, with certain 
restrictions on the range of coupling constants, the second-order fluctuations 
in dS background were all stable on-shell with the exception of the five zero 
modes in the conformal sector.  Although he believed them to be accidental and 
destabilizing beyond one-loop, the arguments in this paper suggest that they 
are a consequence of dS background and will persist to all orders in 
perturbation theory, at least for on-shell quantities.

In Euclidean quantum gravity, the dS background is regarded as the 
sphere\footnote{If one assumes less than maximal symmetry, other 
topologies can also be entertained, but, as discussed in \secn{sec:intro},
non-isometric conformal solutions do not exist except for $S^4.$} $S^4$. 
Assuming the background is $SO(5)$ invariant, it is useful to expand 
the field $h_\ab$ in representations of $SO(5)$ because it diagonalizes the 
operators $\widetilde{\ocal}^{\ge\ab}$ (for a judicious choice of gauge-fixing).
Of course, since the full isometry is not manifested by any choice of 
coordinates, normally an investigation of the Killing equations must be carried out:
\beq\label{eq:killing}
\nabla_a \xi_b+\nabla_b \xi_a=\nabla{\cdot}\xi\,\frac{g_{ab}}{2}.
\eeq
If $\xi^a$ is closed, i.e., $\nabla{\cdot}\xi=0,$ the Killing vector field $\xi^a(x)$ is 
a generator of an isometry of the metric.  (On a contractible manifold, such as 
$S^4,$ a closed vector is exact $\xi_a=\nabla_a X$ for some function $X.$)  If  
$\nabla{\cdot}\xi\neq0,$ then $\xi^a$ is a non-isometric, conformal 
Killing field or a homothetic field. As the name suggests, it is associated with a 
conformal transformation of the metric.  

A more intuitive approach to dS isometries is to embed $S^4$ into flat 
Euclidean spacetime in five dimensions, $E^5,$ where the metric, $\delta_{ij},$  
is trivial in Cartesian coordinates, and the $SO(5)$ isometry is manifest. The 
expansion of the fluctuations in irreducible representations (irreps) of $SO(5)$ is 
far more easily performed in terms of tensors on the co-tangent bundle on 
$E^5$ rather than on $S^4.$ In the next section, we shall review classical dS 
space as a submanifold in five-dimensional flat space.  

\section{de~Sitter space as a submanifold in five dimensions}
\label{sec:ds5}

As mentioned earlier, for the time being, we shall work with Euclidean signature 
and treat classical dS space as the sphere $S^4.$  Later, in 
\secn{sec:lorentz}, we shall comment on what changes are required for 
Lorentzian signature.  We begin by describing $S^4$ as an embedding in flat, 
five dimensional space $E^5$ which, in Cartesian coordinates has the trivial metric:
\beq\label{eq:cartesian}
ds^2=\delta_{ij}dx^idx^j.
\eeq
$S^4$ may be defined 
as the set of all points $x^i$ in $E^5$ satisfying the equation 
\beq\label{eq:submanifold}
\delta_{ij}x^ix^j\equiv \norm{x}^2=r_0^2,
\eeq 
where $r_0$ is related to the on-shell value $R_0\equiv 4\Lambda$ of the 
scalar curvature by $r_0\equiv \sqrt{12/R_0\,}.$ 
It must be shown that this embedding actually corresponds to the dS metric on 
$S^4,$ but we shall take that as given. By 
inclusion\footnote{Our notation and some basic concepts in differential 
geometry are summarized in Appendix \ref{sec:dg}.}, every point on $S^4$ 
can be assigned coordinates in $E^5,$ so one has a mapping from $S^4$ to 
$E^5.$ The co-tangent space on $S^4$ may be regarded as the pull-back of 
the co-tangent space on $E^5.$  
 
The beauty of this description is that it does not require
the explicit introduction of coordinates on $S^4$, so that the isometries of dS 
space are transparent.  That is useful because no single coordinate
system covers all of $S^4,$ and the selection of any particular
coordinate system only reveals a subset of the isometries of the
$S^4$ submanifold. For example, much of $S^4$ is covered by spherical
coordinates, delineated explicitly in \eqn{eq:metricS4}, but the only
evident isometry is the independence of the metric on the angle $\theta_4.$  In fact,
$S^4$ is invariant under global $SO(5),$ as is easily seen,
since the submanifold of $E^5$ is completely  specified by
\eqn{eq:submanifold}.

The embedding of $S^4$ in $E^5$ is a purely geometrical construction in order to 
clarify the isometries of the background field.  It has no effect on the dynamics, 
which always takes place in four-dimensions. Nevertheless, we can describe the 
decomposition of the fluctuations into irreps of $SO(5)$ 
more clearly in this manner. To understand why, we need to spell out the nature of 
the calculation of the Euclidean path integral (EPI). This is a rather long detour 
into explaining the decomposition of fluctuations of the metric in dS space into 
harmonics of $SO(5).$  As a bonus, however, we shall finally understand the 
origin of the five zero modes.  

The construction to follow expands somewhat on discussions given previously 
in \reference{Gibbons:1978ji, Fradkin:1983qk, Avramidi:1986mj}. 
Although $h_\ab(x)$ is a fluctuation of the metric, the point $x$ always remains on 
the background manifold. At every point $x$ on the $S^4$ manifold, the tangent 
space $TS_x^4$ is defined and a set of basis vectors chosen. It can be either a 
coordinate basis $\pa_\alpha$ or a linearly independent tetrad or vierbein 
$e_a(x)=e_a^\alpha(x)\pa_\alpha,$ conventionally taken to be orthonormal in the sense that 
\beq
e^\alpha_a(x)e^\beta_b(x)g^B_\ab(x)=\delta_{ab},\ \ \mathrm{and,\ inversely,}\ 
e_\alpha^a(x)e_\beta^b(x)\delta_{ab}=g^B_\ab(x),
\eeq
where $g^B_\ab(x)$ is the metric on the background $S^4.$
The components of a tangent vector $v$ can be defined either by 
$v=v^\alpha\pa_\alpha$ or, alternatively, by $v=v^a e_a,$ with the relationship 
 between the two descriptions given by the invertible matrix 
$e_a^\alpha(x),$ i.e., $v^\alpha=v^a e_a^\alpha.$ Similarly, 
$dx^\alpha=e^\alpha_a (x)dx^a,$ so that an arbitrary one-form $dv$ may be expanded 
either way, with coordinates related by $dv_\alpha =dv_a e_\alpha^a(x).$
The collection of all such vectors $\mathbf{v}$ is called the tangent fiber at the point $x,$
which is a four-dimensional vector space $TS^4_x.$  Each point $x$ is 
associated with a different tangent space, so the elements of each tangent 
space requires pairing the point $x$ together with the coordinates of 
vectors of the fiber. The collection of all such points with their associated 
fibers, for all possible choice of coordinates, forms the tangent bundle 
$TS^4.$ It is eight-dimensional, requiring four coordinates to specify the point, 
and another four to label the coordinates of each tangent vector. 

Similarly, the co-tangent bundle $T{}^*\!S^4$  has fibers consisting of the 
cotangents $dv$ expressed in either form. The background metric 
$g_\ab^{dS}(x)$ on $S^4$ and the fluctuations $h_\ab(x)$ are the 
components of symmetric, bi-linear functionals defined on the co-tangent 
space at $x$, e.g., in a coordinate basis, $h_\ab(x) dx^\alpha dx^\beta.$  The 
EPI integrates over all the fluctuations $h_\ab(x)$ at each point as well as 
over all points on $S^4.$

One may perform a similar construction on $E^5,$ forming the tangent $TE^5$ 
and co-tangent $T^*\!E^5$ bundles.  Since it is five-dimensional, we must 
introduce a f\"unfbein basis $e_a(x)\equiv e_a^i(x)\pa_i, \{a=1,\ldots,5\},$ 
which is trivial in Cartesian coordinates $e_a^i \equiv \delta^i_a.$ The 
essential difference between the vierbein basis $e_a(x)$ of $TS^4$ and the 
f\"unfbein basis for $E^5$ is of course the radial vector, represented in Cartesian 
coordinates by $e^i_r(x)\equiv x^i/r \equiv \hat{x}^i,$ where $r=\norm{x}.$  Unlike 
\eqn{eq:submanifold}, here $r$ need not be confined to the original $S^4$ 
submanifold with $r=r_0.$  (It may be useful to keep in mind spherical coordinates, 
explicitly given in \eqn{eq:spherical}, for which the metric takes the form of 
\eqn{eq:sphmetric}.)

When we say that the metric on $S^4$ is the pull-back of the metric on $E^5$, 
what we mean is that, since the co-tangent bundle $T{}^*\!S^4$ is the 
pull-back of $T{}^*\!E^5,$ covariant tensors of the latter, such as the metric or 
$p$-forms, may be associated 
with covariant tensors of the former. Since $h_\ab(x)$ is arbitrary, the 
co-tangent bundle $T{}^*\!E^5$ will not 
be $SO(5)$-invariant, but this pull-back, a linear transformation, can be carried 
out regardless because the arguments of these tensor fields lie on $E^5$ and 
$S^4.$ For example, in spherical coordinates, the $E^5$ metric 
takes the form given in \eqn{eq:sphmetric}. To pull back to the $S^4$ manifold,
one may simply fix $r=r_0$ in \eqn{eq:sphE5} and drop the $dr$ term to get 
$ds_4^2=r_0^2d\omega_4^2.$ The dimensionless quantity $d\omega_4^2,$ 
\eqn{eq:metricS4}, denotes the metric on the unit $S^4$ sphere in these coordinates.

Since one can always associate an external normal at a point on an orientable 
manifold, it is worth asking whether some geometrical feature of fluctuations on 
$S^4$ can be associated with the radial direction in $E^5.$ A hint can be found 
in the Hodge-dual or Hodge-star of the one-form $e^r(x)\equiv e^r_i(x) dx^i:$ 
\beq\label{eq:hodge}
{}^{\mathbf\star}e^r(x)=
e^{a_1}(x)\wedge e^{a_2}(x)\wedge e^{a_3}(x)\wedge e^{a_4}(x),
\eeq
 where the $e^{a_j}(x)$ are an (appropriately ordered) orthonormal basis of 
 $T^*\!S^4$ at the point $x.$  The right-hand side of \eqn{eq:hodge} is 
 proportional to the coordinate-invariant volume co-form on $S^4$:
\beq\label{eq:dv4}
dV_4\equiv \sqrt{g_4}\,
e^{a_1}(x)\wedge e^{a_2}(x)\wedge e^{a_3}(x)\wedge e^{a_4}(x).
\eeq
Thus, up to the factor of $\sqrt{g_4}\,,$ we may identify the radial direction 
in $E^5$ with $dV_4.$  

Another way to view this relation is to start from the volume co-form in $E^5$:
\beq\label{eq:dv5}
dV_5\equiv \sqrt{g_5}\,
e^r(x)\wedge e^{a_1}(x) \wedge e^{a_2}(x) \wedge e^{a_3}(x) \wedge e^{a_4}(x) .
\eeq
The relations above can be conveniently stated in a coordinate independent 
fashion in terms of the interior product $\iota_v,$ which maps forms of order 
$p$ to forms of order $p{-}1,$ such as the mapping from $dV_5$ to $dV_4.$ 
(See Appendix~\ref{sec:dg}.)  
In this language, the unit area of $S^4$ at a point $x$ is given by the four-form 
\beq\label{eq:ier}
dV_4={\bf \iota}_{e_r}[dV_5].
\eeq
Correspondingly, a radial vector $v(x) e_r(x)$ at a point $x$ is dual to a
rescaled local volume element $v(x) dV_4$ on $S^4.$ (N.B. $g_4 
\neq g_5|_{r=r_0}.$)

The upshot of this is that the differential ``surface area" $dV_4$ on $S^4$ may 
be associated with the contraction of the five-form volume $dV_5$ with the unit 
normal $\hat{e}_r(x)$ in $TE^5,$ evaluated on the submanifold $r=r_0.$ If the 
$S^4$ metric fluctuates in a way that changes its surface area  
(volume form on $S^4$) by some amount, it can equivalently be expressed as a 
certain \emph{rescaling} of the magnitude of the normal vector in $E^5.$  
This relationship is a key to understanding how conformal 
fluctuations of the volume form on $T{}^*\!S^4$ are related to radial 
fluctuations on $T{}^*\!E^5.$

\section{Lifting metric fluctuations from $S^4$ to $E^5.$}
\label{sec:lift}

Now we wish to consider metric fluctuations $h_\ab$ on $T{}^*\!S^4$ and, in 
particular, to spell out how they are reflected in $T{}^*\!E^5.$  To simplify the 
discussion, We shall suppress gauge-dependent fluctuations and work with 
those that survive on-shell. For this purpose, it is helpful to choose a unitary 
gauge in which the gauge degrees of freedom vanish, such as the 
transverse-traceless (TT) gauge in which 
\beq\label{eq:ttgauge}
h_\ab=\frac{h}{4}g^B_\ab+h_\ab^\perp,
\eeq
where $h\equiv g^B{}^\ab h_\ab$ and $\nabla^B{}^\alpha h_\ab^\perp=0,$ setting 
the other four gauge-dependent modes to zero\footnote{This is not the most 
convenient choice for understanding renormalizability, but we are at present only 
concerned with the issue of stability.}. The fluctuations $h_\ab^\perp(x)$ form a 
symmetric traceless tensor on the co-tangent space $T{}^*S_x^4$ that vanish 
when evaluated on the normal $e_r(x),$ which can easily be visualized 
from the embedding of $TS^4$ in $TE^5.$ Thus, they do not change the volume 
element $dV_4$ at $x.$ On the other hand, the conformal fluctuations 
\beq\label{eq:conformal}
g^B_\ab(x)\left(1+\frac{h(x)}{4}\right)
\eeq
change the volume form $dV_4$ through its rescaling of $\sqrt{g_4(x)}$ by 
$(1+h(x)/4)^2.$ As discussed in the preceding section, via the Hodge dual, this 
can be pictured as a change in the radial component $g_{rr}dr^2$ of the metric on 
$T{}^*\!E^5$:
\beq\label{eq:ds5}
ds_5^2=(1+h(x)/4)^2dr^2+r^2d\omega^2_4,
\eeq 
where the metric is to be evaluated at $r=r_0.$  
In this rather round-about way, we have lifted the conformal 
fluctuations of the metric on $T{}^*\!S^4$ to fluctuations of the metric on the 
co-tangent bundle $T^*\!E^5.$ 

To extend this to include the fluctuations $h^\perp_\ab$ is straightforward since 
these are tensors built on the co-tangent bundle $T{}^*\!S^4,$ which is  
represented in $E^5$ by co-vectors normal to $e^r(x^\mu)$ for all 
$x^\mu.$ Thus, fluctuations confined to $T{}^*\!S^4$ are unchanged in passing to 
$T{}^*\!E^5.$ Altogether then, in the TT-gauge, \eqn{eq:ttgauge}, 
the fluctuations of the metric in $T{}^*\!S^4$ may be 
represented by fluctuations of the metric of $T{}^*\!E^5$ 
implicitly defined in spherical coordinates by  
\beq\label{eq:invlngth}
ds_5^2=(1+h(x)/4)^2dr^2+\left(g^B_\ab 
+h^\perp_\ab\right)dx^\alpha dx^\beta,
\eeq
with the understanding that $\alpha,\beta$ refer to coordinates on $TS^4$ or 
$T^*\!S^4.$ On-shell, we must set $r=r_0.$  

As can easily be seen in Cartesian coordinates, the symmetries of $E^5$ are 
the Poincar\'e semigroup $SO(5)\rtimes P^5,$ where $P^5$ represents translations.  
The generators of translations, $P^5,$ commute with each other but 
transform as a vector $\bf 5$ under $SO(5).$ These isometries may also be 
inferred from the Killing equations, \eqn{eq:killing}, which, in $E^5,$ become
\beq\label{eq:kill}
\nabla_i\xi_j+\nabla_j \xi_i=0.
\eeq
In Cartesian coordinates, where the spin connection vanishes, the general 
solution is $\xi^i=\omega^i_{\; j} x^j+k^i$, for constants $k^i$ and antisymmetric 
matrix $\omega^{ij}.$  The first term represents the 10 generators of 
$SO(5)$ rotations; the second term, of 5 translations. Since 
$x^i\omega_{ij}x^j=0,$ the rotation generators $\omega^i_{\; j} x^j$ are 
orthogonal to the radial vector $x^i;$ they lie within the tangent bundle $TS^4.$ 

In going on-shell, $r=r_0,$ the translation symmetry is 
broken but $SO(5)$ is preserved.  We shall return to the consequences of the 
breaking in the next section, but a major benefit of the lifting of fluctuations from 
$S^4$ to $E^5$ is that the decomposition of $h(x)$ in terms of irreducible 
representations (irreps) of $SO(5)$ is far simpler in flat 
five-dimensional $E^5$ than on four-dimensional $S^4.$   As discussed in 
Appendix~\ref{sec:sphh}, the basis functions for these irreps correspond to 
harmonics $f_n(x)$ on $E^5$ that depend on the components of the single 
vector $x^i\pa_i.$  In Cartesian coordinates, each harmonic is a polynomial 
of degree $n$ satisfying Laplace's equation 
$\Box_5 f_n=\sum_i \pa^2 f_n/\pa x^i{}^2=0.$  The irreps $f_n(x)$ are given by 
the symmetric, traceless tensors $\scal^{ijk\ldots}(x)$ of degree $n.$ In spherical 
coordinates on $E^5,$ they take the form $f_n^m=r^n \phi_n^m(\omega^\alpha),$ 
where we added an index $m$ that runs over the number of linearly 
independent tensors of degree $n.$ The $\phi_n^m(\omega^\alpha)$ are 
spherical harmonics on the unit $S^4,$ which, conventionally, are taken to be 
an orthonormal basis of functions.  Any nonsingular scalar field on $S^4,$ such as 
$h(x),$ may be expanded as $h(x)=\sum h_n^m \phi_n^m(\omega^\alpha).$
 
Some further details of these representations are given in 
Appendix~\ref{sec:sphh} and have been reviewed in \reference{Gibbons:1978ji}.
Suffice to say that the five zero modes of particular interest to us correspond to 
$n=1,$ having $\phi_1^m(\omega^\alpha)=x^m/r_0,$ $\{m=1,\ldots,5\}.$ The 
$\phi_1^m$ can be taken to be proportional to the five 
functions $x^m/r$ in \eqn{eq:spherical}. The question is why these particular 
modes $\sum h_1^m\phi_1^m(\omega^\alpha)$ turn out to be zero modes. 

\section{The five zero mode fluctuations}
\label{sec:5zero}

In the embedding $S^4\to E^5,$ we assigned $S^4$ to the submanifold 
$\delta_{ij}x^ix^j=r_0^2,$ but we could equally well have chosen the submanifold
\beq\label{eq:s4b} 
\delta_{ij}(x^i{-}b^i)(x^j{-}b^j)=r_0^2,
\eeq
for some constant five-vector $b^i.$  From the point of view of $E^5,$ this 
corresponds to the $S^4$ sphere centered at $b^i$ rather than at the origin; 
call it $S^4_{\bf b}.$ The classical action will be the same on $S^4_{\bf b}$ 
as on $S^4_{\bf 0}$ because the two manifolds are diffeomorphic.  This can 
be explicitly seen, for example, by replacing $x^i$ by $x^i-b^i$ in the definition 
of spherical coordinates, \eqn{eq:spherical}.  The metric on $E^5,$ 
\eqn{eq:cartesian}, is invariant under translations, so the projection onto 
$S^4_{\bf b}$ will give the same functional form $g_\mn(\omega^\alpha)$ in 
spherical coordinates as its projection onto the original sphere $S^4_{\bf 0}.$  
Thus, $b^i$ are moduli characterizing the embedding in $E^5$ but having no 
physical relevance to the dynamics on $S^4_{\bf b}.$

For infinitesimal $\Delta b^i,$ the first-order change in the submanifold 
$S^4_{\bf 0}$ is $x^2-2\Delta{b_i}x^i=r_0^2,$ so locally, the change appears to be 
a radial displacement or, better, a rescaling\footnote{This scaling may be seen 
embedding $S^4$ into the light-cone in six-dimensional Minkowski 
space $M^{5,1},$ for which the isometries are the conformal group, $SO(5,1).$  
See, e.g., appx.~D of \reference{Kosyakov:2007qc}.} of the radius by 
$(1+\Delta{b^r}/r_0).$  Clearly, a change of embedding from $b^i$ to 
$b^i+\Delta b^i$ in \eqn{eq:s4b} is identical to a change $x^i{\to}x^i-\Delta b^i$ at 
every point on $S^4_{\bf b}.$ So, instead of regarding this as a change of $b^i,$ 
we may instead think of it as a change of $x^i$ or, in the QFT, for fixed $x^i,$ a 
fluctuation in the metric of the form $(1+h/4)^2dr^2.$ It is equivalent to an 
infinitesimal change\footnote{In the notation used earlier and in 
Appendix~\ref{sec:sphh}, $\hat{x}^m=\phi_1^m(\omega^\alpha),$ the five 
spherical harmonics.} 
$\Delta h=-4\Delta{b_i}\hat{x}^i/r_0.$  This is precisely the form of the $SO(5)$ 
vector fluctuation $\Delta{h_{\bf5}}=\Delta{h_i}\hat{x}^i,$ with 
$\Delta{h_i}=-4\Delta{b_i}/r_0.$  

This last statement, which agrees with the intuitive idea that an infinitesimal 
translation in some direction involves both radial and tangential displacements, is 
worth expanding upon.  Even though locally, the rescaling is radial, it must of course 
be true for every point on $S^4.$ Consequently, if $v_r(\omega^\mu)=0$ for all 
points on $S^4,$ then\footnote{This argument breaks down where our spherical 
coordinates become singular, but one can patch these to another set of spherical 
coordinates with the axes rotated.} $v_i=0$ on $S^4$ for all $i$. Since 
$v_r \equiv v_i e^i_r$ and $e^i_\alpha=r_0\,\pa_\alpha e^i_r,$ \eqn{eq:funfbein}, 
it follows that 
$\pa_\alpha v_r=(\pa_\alpha v_i)e^i_r+(v_i/r_0) e^i_\alpha=0.$ 
Since $e^i_r$ and $e^i_\alpha$ are orthogonal, each term must vanish separately. 
Hence, $v_\alpha=0$ for all angles $\omega^\alpha$, so 
$\mathbf{v}=0$ on $S^4$ in any frame. 

Therefore, this particular coherent fluctuation is equivalent to an infinitesimal 
displacement of the entire manifold, under which, according to the preceding 
analysis, the action is invariant.  This is \emph{why} these five vector modes in 
$h_{\bf5}$ are zero modes. In terms of the metric on $S^4,$ \eqn{eq:ttgauge}, we 
know that these correspond to conformal fluctuations. Indeed, we show in 
Appendix~\ref{sec:kill} that these infinitesimal translations also correspond to 
conformal Killing fields, which are not usually associated with symmetries of the 
action. We also show that the conformal Killing fields can be associated locally with 
$SO(5)$ rotations, so they in fact do generate zero modes.

If we regard the $S^4_{\bf b}$ sphere as having a fixed center $\bf b$ when 
embedded in flat five-dimensional space, then these zero modes would not be 
allowed fluctuations of the metric. This strongly suggests that these coherent 
fluctuations associated with $h_{\bf5}$ are unphysical.

The preceding arguments do not depend upon the quadratic approximation to 
the fluctuations, and, since it is based on the underlying dS symmetries, it is plausible 
that the argument would extend to all orders in perturbation theory for the effective 
action, \eqn{eq:effaction}. These zero modes are a consequence of the maximal 
global symmetry assumed for the background. 

Of course, if this background were unstable, then the assumption of global 
$SO(5)$ symmetry would become questionable. At the least, it is necessary that, at a 
local extremum, the quadratic approximation should have no negative eigenvalues. It 
is well-known that the fluctuations in E-H gravity do have negative eigenvalues, so 
whether or not there are also zero modes is somewhat of a moot point, but these 
same zero modes do occur there.  For renormalizable gravity, 
Avramidi~\cite{Avramidi:2000bm, Avramidi:1986mj} observed that, for a range of the 
couplings $a,b$ there were in fact no unstable modes and that the only zero modes 
were the five discussed here.  In contrast to his conclusion, however, we do not 
expect these zero modes to be removed in higher order, so it remains to be 
explained how they should be handled.

If we do not integrate out the fluctuations, the second-order calculation constitutes a 
test of whether the Euclidean action is classically stable. To the extent that we may 
regard rigid translations in five-dimensions of the dS submanifold as unphysical, these 
coherent zero modes do not in fact constitute a flat direction in the physically allowed 
space of fluctuations. In that case, we may conclude that the Euclidean classical 
EOM are in fact stable for a range of the couplings.

Returning to the QFT,  one integrates over the second-order fluctuations 
to obtain the one-loop corrections to the classical action.  To make use 
of such a calculation requires going off-shell in order to be able to take variational 
derivatives to determine corrections to the EOM and correlation functions. 
Such calculations are inherently gauge-dependent, although the value of the 
action on-shell is not.  Further, the EPI off-shell cannot be analytically performed 
except in cases when the background field is assumed to have a high degree of 
symmetry. Assuming the background field retains maximal symmetry, only the 
scalar curvature $R$ needs to be determined. (This calculation is analogous to 
calculating the effective potential in ordinary, flat-space field theory.) It has been 
carried out~\cite{Avramidi:2000bm, Avramidi:1986mj} for a range of 
gauges. Off-shell, there appear to be no zero modes, at least for some gauge 
choices, and the one-loop correction can be carried out in a neighborhood of the 
classical curvature. The vanishing of the first variation then determines the 
one-loop correction to $R$. 

Christensen and Duff~\cite{Christensen:1979iy} calculated the value of the 
one-loop corrections to the effective action in E-H gravity assuming an 
\emph{arbitrary} background field but using the classical EOM, 
$R_\mn=\Lambda g_\mn$ in order to obtain a gauge-invariant result. (A 
generic background field presumably has no zero modes.) They also 
performed the calculation in a maximally 
symmetric background and showed that agreement with their first calculation, 
when restricted to dS background, is obtained only if the zero 
modes, including the five non-isometric conformal modes, are properly 
accounted for. We do not doubt this conclusion; these zero modes are certainly 
present, even for the E-H effective field theory, but their interpretation is a matter 
for further discussion. As we have argued, they are a legacy of a symmetry between 
distinct but identical spacetimes in $E^5.$

Unlike more familiar applications~\cite{Gervais:1975yg}, including the one involving 
the partition function for Schwarzschild black holes treated in Sec.~3 of 
\reference{Gibbons:1978ji}, the collective coordinates $b^i$ associated with these 
coherent fluctuations, analogous to their $q^m,$ are not dynamical coordinates  
associated with metric fluctuations and do not reflect a physically realizable flat 
direction. 

\section{Some comments on Lorentzian signature}\label{sec:lorentz}

Although we shall leave the case of Lorentzian signature for future work, we 
shall indicate some of the differences and challenges that occur and offer some 
conjectures about what we expect to find. There is no problem starting from the 
classical solution without choosing coordinates.  Embedded in five-dimensional 
Minkowski space, the dS solution corresponds to the hyperbolic submanifold 
$H^4$ described by $x^ix^j\eta_{ij}=r_0^2,$ where 
$\eta_{ij}=\rm{Diag}\{1,1,1,1,-1\}.$ The Minkowski metric $ds^2=\eta_{ij}dx^idx^j,$ 
has global isometries associated with the Poincar\'e semigroup, 
$SO(4,1)\rtimes P^5,$ which is broken to $SO(4,1)$ on the hyperboloid $H^4.$  
The hyperboloid has topology $R{\times}S^{n-1};$ the main difference from the 
Euclidean case is that the manifold is no longer compact.  

As before, we can discuss the diffeomorphic family of hyperboloids 
$H^4_{\bf b}$ defined by $(x^i-b^i)(x^j-b^j)\eta_{ij}=r_0^2,$ for fixed vector $b^i.$ 
There will no doubt be zero modes of the fluctuations of each manifold associated 
with infinitesimal changes of $b^i.$  We would expect that their treatment should 
be analogous to the Euclidean case; they represent unphysical fluctuations.  

There are two things one would like to investigate for the Lorentzian case, viz., 
unitarity and the possible role of Euclidean 
instantons\cite{Strominger:1984zy, Fradkin:1983qk}.  Concerning unitarity, the 
definition of the Hilbert space of states and the associated norm is frame 
dependent already in a curved background, even before quantizing gravity. (See, 
e.g., \reference{Birrell:1982ix, Parker:2009uva}.)  The definition of the Hilbert 
space of states is usually associated with fixed time slices (spacelike submanifolds 
with timelike normal,) but there are a great many possibilities because the global 
symmetry is maximal. The natural choice for discussing unitarity in a Hamiltonian 
framework would be static coordinates:
\beq\label{eq:static}
ds^2=-\left(1-\frac{\rho^2}{r_0^2}\right)dt^2+
\left(1-\frac{\rho^2}{r_0^2}\right)^{-1}d\rho^2+\rho^2d\omega_2^2,
\eeq
for $0\leq\rho<r_0.$
These coordinates obviously develop singularities at the 
cosmological horizon $\rho=r_0$ and therefore only cover a portion of the dS 
manifold. This situation is very much like the horizon for the Schwarzschild 
black hole in static coordinates, which can only be reached asymptotically, but this 
is a property of the frame and not a singularity of the manifold. Unlike the BH, one 
can find other frames that do cover the entire manifold without encountering a 
true spacetime singularity.  For example, in global coordinates, 
\beq\label{eq:global}
ds^2=-dt^2+r_0^2\cosh^2(t/r_0)d\omega_3^2,
\eeq
in which the isometry $R{\times}S^3$ is manifest.  The existence of global 
coordinates that cover the entire manifold is another difference from the 
Euclidean case. Normalizability at fixed time $t$ once again reduces to 
normalization on a compact manifold $S^3$. How to demonstrate unitarity 
in such time-dependent backgrounds is unclear.  Further, the usual definition of a 
well-defined no-particle state (``vacuum") includes cluster decomposition of 
correlation functions, something that appears to be impossible on a compact manifold.

On the other hand, in a semi-classical approximation, it may be possible to 
understand the effects of the instanton as a tunneling amplitude between $SO(4)$ 
coverings of $S^3$ in the distant past with coverings in the distant future.  If so, 
the Gauss-Bonnet coupling constant may acquire a dynamical significance 
analogous to the $\varTheta$ parameter in QCD.  Since 
$SO(4){\cong}SU(2){\otimes}SU(2),$ the parallels may be very close, 
as the topology of $SU(2)$ is\footnote{The two $SU(2)$'s may be called left 
and right, and there may well be chiral representations related to the Hirzebruch 
signature, and instantons associated with this index~\cite{Strominger:1984zy} as 
well.} $S^3.$   We hope to return to these questions in the future.

\section{Discussion}\label{sec:discuss}

Zero modes of scalar fields in curved spacetime have been discussed for a 
long time. (See, e.g., \reference{Birrell:1982ix, Parker:2009uva}.) One lesson 
learned is that free massless scalars can sometimes be misleading and ought 
to be examined as limits of interacting QFTs or in the context of dynamical 
gravity.  Minimal coupling (absence of a $\phi^2 R$ interaction), for example, 
is not a fixed point of any nonsupersymmetric, interacting theory, so such 
models~\cite{Folacci:1992xc} should be examined with 
care~\cite{Einhorn:2002nu}.

Folacci~\cite{Folacci:1996dv} has considered a simplified version of the 
problem of the five zero modes in dS background\footnote{See also the 
appendix to \reference{Folacci:1992xc}.}. He associates the zero 
modes with a ``five-dimensional gauge transformation" 
$h(x) {\to} h(x)+h_{\bf5}.$ The effective action depends only upon the 
background field, $g_\ab^B$ not on $h(x),$ but it is certainly true that, to one 
loop order, if $h_{\bf5}$ is expressed in terms of the five spherical harmonics on 
$S^4$, the on-shell effective action does have the symmetry 
$\Gamma[g_\ab^B]=\Gamma[g_\ab^B(1+h_{\bf5}(r_0,\vartheta_\alpha))^2].$ (See 
\eqn{eq:invlngth}.)   A gauge-symmetry is designed to remove an unphysical 
degree of freedom in local field theory, but this local symmetry is a remnant 
of the global symmetries and would not be present in another background.
As discussed in the Introduction, \secn{sec:intro}, these 
modes are peculiar to spacetimes with maximal global symmetry.  Because 
these are non-isometric zero modes, we cannot even associate a conserved 
current with them. It is hard to see how to associate this property with a 
gauge symmetry of the action on $S^4.$  

Folacci~\cite{Folacci:1996dv} also suggests that the zero modes are a 
consequence of the compactness of the sphere $S^4$ of Euclidean dS and 
would not be present for the non-compact hyperboloid $H^4$ of Lorentzian 
dS. Consequently, he argues, they present a barrier to analytic 
continuation from Euclidean to Lorentzian signature. It is true that on a 
compact manifold, modes may be allowable whose Lorentzian analogs would 
be non-normalizable because the spacetime volume becomes infinite.   However, 
as indicated in the preceding section, the normalization of states
should be performed on fixed time-slices, for which the metric is compact 
(except for the so-called Poincar\'e slice).  Further, we can explicitly turn 
spherical coordinates on $S^4,$ \eqn{eq:spherical}, into global coordinates on 
$H^4$, \eqn{eq:global}, by replacing $\vartheta_1{\to} \pi/2-it.$  There is no 
difficulty normalizing the conformal Killing modes at fixed $t.$ 

Follacci~\cite{Folacci:1996dv} further argues that the S-matrix will be 
infrared-divergent.  Even if there were an S-matrix, the same may be said 
about quantum electrodynamics (QED) in Minkowski space, for which the true 
asymptotic states of the theory are orthogonal to the Fock states. This ``IR catastrophe" 
is not insurmountable~\cite{Bloch:1937pw}. The asymptotic states non-relativistically 
are Coulomb wave functions and, relativistically, are probably coherent 
states~\cite{Kulish:1970ut, Contopanagos:1991yb}.  In any case, these IR 
divergences do not prevent QED from making contact with the real world. The infinity 
of Fock states must be summed up to form observables having a finite energy 
resolution. The IR divergences in gravity are no worse than in 
QED~\cite{Weinberg:1965nx} (at least not in asymptotically flat spacetimes).  
So, even though Folacci's arguments may be 
formally correct, it must be shown that the IR divergences prevent 
predictions analogous to those of QED, when rephrased in terms of the  
limits of hypothetical measurements of limited accuracy.  Since there is no 
S-matrix when $\Lambda\ne0,$ we must study long-time, long-distance correlation 
functions on-shell.  Further work is needed to determine just what the infrared 
sensitivity of dS spacetime implies and how Lorentzian correlation functions 
may be related to their Euclidean counterparts.

Although it is often said that there are no local observables in gravity, in reality, all 
measurements are determined by the apparatus used. Theory may be used to 
relate them to the distant past (e.g., in astrophysics and cosmology) but both 
observationally and theoretically, calculations involve gauge-invariant correlations 
over finite times and distances.  The actual measurements single out special 
``frames" and, from the point of view of the path integral, select a particular set of 
histories\footnote{See, e.g., Gell-Mann \& Hartle\cite{Gell-Mann:2013hza} and 
references therein.\label{ftnt:gmh}}. Such correlation functions will not be IR 
divergent, but, to complete the story, one must investigate the character of the 
dependence on large distances and long times in order to establish that 
observables in dS spacetime can be expressed in terms of the accuracy of 
the measuring apparatus. 

\section{Conclusions}\label{sec:conclude}

Clearly, there is more that must be done to clarify these infrared issues for 
Lorentzian signature, but these matters seem to be essentially unrelated to the zero 
modes of interest here. More generally, the nature of measurement introduces 
apparatus that selects nearly classical histories of one sort or another, so that 
spacetime events decohere. (See footnote~\ref{ftnt:gmh}.) This will necessarily break 
exact dS invariance, so this discussion may be delicate but hopefully will be 
controllable in a manner similar to QED. 

It has sometimes been suggested\footnote{See, e.g., 
\reference{Avramidi:2000bm} and the summary of earlier literature in 
\reference{Folacci:1996dv}.} that these modes are a feature at one-loop 
and unlikely to be sustained in higher-order.  Since we have a renormalizable 
theory of gravity, we ought to be able to answer this, at least in 
principle. Nobody has done calculations beyond one-loop order, but our 
arguments in \secn{sec:5zero} depend only upon the symmetries of the 
background and not on the order in the loop expansion.  Unless some unstable  
modes arise in higher-order, our conclusions should be good to all orders.
Since the renormalizable theory is asymptotically free, in fact, the one-loop 
approximation ought to be good at high scales, so the absence of negative modes 
should not be undermined by higher-order corrections.

Analogous phenomena may occur in other models in which the 
background (or condensate) is assumed to have certain continuous global 
isometries. Simply subtracting zero modes should not be done without 
understanding their origin.  

The assumption of exact dS symmetry is not correct in any relevant 
cosmological application since the presence of matter will lead to a 
stress-energy tensor that will contribute a background energy density and 
pressure for which the equation of state differs from that associated with a 
cosmological constant.  Thus, more realistic cosmologies (such as the 
$\Lambda$-Cold Dark Matter model or, more generally, a 
Friedmann-Lema\^itre-Robertson-Walker metric) will break dS symmetry, and 
these non-isometric zero modes will disappear since they depend crucially upon 
the assumption that the background topology is $S^4$ or $H^4.$  

There are other reasons to doubt that dS is the correct background in any realistic 
cosmology. The dS metric assumes that the isometries are eternal, and it is unlikely 
that its symmetry between past and future is correct for applications to our universe. 
For example, inflationary cosmologies suggest that the dS approximation is 
only good for a finite period of time, having both an initial time when the exponential 
expansion begins and a final time when it effectively ends. For example, 
assuming that there is a finite initial time in the distant past only after which the dS 
metric becomes a good approximation is already a significant modification, which 
is sufficient to cure some infrared 
problems~\cite{Dolgov:1994ra, Dolgov:1994cq}.
Whether there is any sense 
to the other times described by the dS metric, for example the period of contraction 
rather than expansion, depends on speculations about the universe before the big 
bang, which may or may not have observable consequences for our universe. 
Most cosmologists, at least those exploring inflation, assume that a better 
approximation to dS spacetime is, during inflationary expansion, to adopt 
Poincar\'e coordinates and take only half of the dS manifold. That seems plausible 
although it would be nice not to identify the approximation with a special 
coordinate frame. 
With the addition of matter, the questions become more complicated since 
other fields may condense.  However, we have seen that these zero modes 
remain present, not only in the models considered by others, but also in all 
the models that we have 
examined~\cite{Einhorn:2014gfa, Einhorn:2015lzy, Einhorn:2016mws}. There 
is every reason to expect that they remain to the extent that the background is 
well approximated by dS space.  

For renormalizable gravity, having argued that these zero modes are unphysical 
and do not represent flat directions, we can conclude that dS 
space is perturbatively stable for some range of couplings.  This will remain correct 
in the presence of matter, at least so 
long as all couplings are asymptotically free~\cite{Einhorn:2016mws}. In a 
separate publication~\cite{Einhorn:tbp}, we shall discuss these matters further 
and explore the spectrum and, to a limited extent, the meaning of unitarity.  

Asymptotically-free, renormalizable gravity can, at worst, be used to suggest 
some new cosmological possibilities, or, at best, to provide a consistent 
extension of quantum gravity within QFT.  Having shown that there are models 
that are asymptotically free in all couplings that do not require 
fine-tuning~\cite{Einhorn:2016mws}, it may even be that renormalizable gravity 
is a consistent completion of Einstein gravity.  Demonstrating unitarity remains 
the outstanding problem.

\acknowledgments
We would like to acknowledge helpful correspondence with I.~Avramidi, M.~Duff, 
and A.~Folacci. One of us (MBE) has benefitted from extensive discussions 
with A.~Vainshtein.  This research was supported in part by the National Science 
Foundation under Grant No. PHY11-25915 (KITP) and by the Baggs bequest 
(Liverpool). 

\appendix

\section{Basic concepts and notation}\label{sec:dg}

Euclidean $dS^4$ can be depicted as the four-sphere $S^4$ of radius 
$r_0=\sqrt{12/R_0},$ where $R_0$ is the on-shell value of the scalar curvature.  
The isometries of $dS^4$ are most easily displayed by embedding $S^4$ in 
$E^5.$  On $E^5$, we imagine setting up five Cartesian coordinate axes. 
Tangents to the coordinate axes form vectors denoted by 
$\pa_i\equiv \pa/\pa x^i,$  which form the basis of a vector space.  Each point 
$x$ in $E^5$ can  be assigned coordinates $x^i$ in $R^5$ according to the 
decomposition\footnote{Concerning nomenclature, $E^5$ without the metric is 
often referred to as $R^5,$ although the notation in the literature is not uniform 
and sometimes these are used interchangeably.} $x=x^i\pa_i.$  The duals to $
\pa_i$ are denoted by $dx^i,$ linear functionals or one-forms on $E^5,$ with 
$dx^i[\pa_j] \equiv \delta^i_j.$  The five $dx^i$ form a basis for the vector space 
$T{}^*\!E^5$ of one-forms, $da=a_i dx^i.$  (One-forms are frequently referred to 
as covariants, in contrast to vectors, which are sometimes called contravariants.)

$E^5$ is equipped with the metric $\delta_{ij},$ the components of the 
symmetric, covariant tensor $ds^2=\delta_{ij}dx^i dx^j,$ where 
$dx^i dx^j$ stands for the direct product $dx^i{\otimes} dx^j$. This may be used 
to define the standard norm $\norm{v}=\sqrt{\delta_{ij} v^i v^j}.$  This implies the 
usual Cartesian inner product of vectors 
$\vec{v}{\cdot}\vec{w}\equiv \delta_{ij}v^iw^j.$  The direct product is to be 
contrasted with the antisymmetric exterior product or two-form 
$\alpha^{ij}=dx^i{\wedge} dx^j=-dx^j{\wedge} dx^i.$ 

The symmetries of $E^5$ are the Poincare group $SO(5){\rtimes}P^5,$ 
a 15-dimensional  group consisting of arbitrary translations of a point together with 
rotations in five dimensions.  Of course, one may go on to discuss other coordinate 
systems on $E^5,$ such as cylindrical, parabolic, elliptic, bipolar, etc. For example, 
spherical coordinates may be defined on $E^5$ by
\begin{align}\label{eq:spherical}
\begin{split}
x^1&=r\cos\vartheta_1,\\
x^2&=r\sin\vartheta_1\cos\vartheta_2,\\
x^3&=r\sin\vartheta_1\sin\vartheta_2\cos\vartheta_3,\\
x^4&=r\sin\vartheta_1\sin\vartheta_2\sin\vartheta_3\cos\vartheta_4,\\
x^5&=r\sin\vartheta_1\sin\vartheta_2\sin\vartheta_3\sin\vartheta_4,
\end{split}
\end{align}
where $0<\vartheta_\alpha<\pi\ \{\alpha=1,2,3\}$, $0\leq \vartheta_4 <2\pi.$
The corresponding metric on $E^5$ then takes the form 
\begin{subequations}\label{eq:sphmetric}
\begin{align}
\label{eq:sphE5}
ds^2&=dr^2+r^2 d\omega_4^2,\\
\label{eq:metricS4}
d\omega_4^2&=d\vartheta_1^2{+}\sin^2\vartheta_1\,d\vartheta_2^2{+}
\sin^2\vartheta_1\sin^2\vartheta_2 \,d\vartheta_3^2{+} 
\sin^2\vartheta_1\sin^2\vartheta_2\sin^2\vartheta_3 \,d\vartheta_4^2.
\end{align}
\end{subequations}
The singular character of coordinates is revealed by the 
vanishing of $g\equiv \det{g_\ab}.$  For these spherical coordinates,
the determinant is
\beq\label{eq:g5}
g_5^{dS}=r^8\sin^6\vartheta_1\sin^4\vartheta_2\sin^2\vartheta_3,
\eeq
which is obviously singular at $r=0$ or when any of these three 
$\vartheta_i=0,\pi.$ This restricts this coordinate patch on $E^5$ 
to exclude these values.

An orthonormal basis in five dimensions in spherical coordinates can be 
chosen as 
\begin{subequations}\label{eq:funfbein}
\begin{align}
\label{eq:funfbeinr}
e^i_r&= 
{x^i}/{r}&&=(c_1,s_1c_2,s_1s_2c_3,s_1s_2s_3c_4,s_1s_2s_3s_4),\ \mathrm{and}\ 
e^i_\alpha,\ \mathrm{where} \hskip0.5in\\
e^i_1&= r\pa_{\vartheta_1}e^i_r&&=
 r(-s_1,c_1c_2,c_1s_2c_3,c_1s_2s_3c_4,c_1s_2s_3s_4),\\
e^i_2&= r\pa_{\vartheta_2}e^i_r&&= r s_1(0,-s_2,c_2c_3,s_2c_3c_4,s_2s_3s_4),\\
e^i_3&= r\pa_{\vartheta_3}e^i_r&&= rs_1s_2(0,0,-s_3,c_3c_4,c_3s_4),\\
e^i_4&= r\pa_{\vartheta_4}e^i_r&&=
 rs_1s_2s_3(0,0,0,-s_4,c_4),
\end{align}
\end{subequations}
where we have abbreviated 
$c_k{\equiv}\cos\vartheta_k, s_k{\equiv}\sin\vartheta_k,$ 
$k=1,\ldots,4.$ These f\"unfbein satisfy $e^i_\mu e^j_\nu \delta_{ij}=g_\mn.$ 
The inverse of the matrix $[e^i_\mu]$ will be written as $[e^\mu_i],$ so it is 
necessary to adhere to our notational conventions using Latin indices for 
Cartesian coordinates and Greek indices for spherical coordinates. (An exception 
is the use of $r$ rather than $\rho$ for the radial coordinate.)

As mentioned in the text, one can replace $x^i$ by $x^i-b^i$ in \eqn{eq:spherical} 
for any constant five-vector $\bf b,$ without making any changes in the metric on
a subdomain at fixed $r_0=\norm{\bf{x-b}}$; consequently, the value of the classical 
action, e.g., \eqn{eq:scl}, on $S^4_{\bf b}$ is independent of $\bf b.$  

As a brief refresher on the Cartan formalism, an exterior differential $d$ takes a 
$p$-form $\alpha$ to a $p{+}1$-form denoted $d\alpha.$ Recall that the 
exterior differential of a function 
$f(x)$ ($0$-form) is the usual differential $df[x]= dx^i \pa_i f,$ discussed above.  
If $\alpha=a_idx^i$ is a one-form, then $d\alpha\equiv da_i{\wedge}
dx^i=(\pa{a_i}/\pa{x^j})dx^j{\wedge}dx^i$ in any coordinates. Similarly, for an 
arbitrary $p$-form. The exterior derivative $d$ has the property that 
$d^2\alpha{=}0$ on any form~$\alpha.$  

The interior product or contraction operator is an operation $\iota_v$ associated 
with a vector $v$ that takes a $p$-form $\alpha$ into a $p{-}1$-form according to 
$\iota_v[\alpha]\equiv \alpha[v,\ldots],$ which symbolically means ``evaluate the 
$p$-form on the vector $v.$"  On a zero-form (i.e, a function) $f(x)$, 
$\iota_v[f]{\equiv}0.$ On a one-form, such as $df,$ $\iota_v[df]\equiv df[v]=v^i\pa_if,$ 
the usual directional derivative.  For a two-form, e.g., 
$\alpha=df{\wedge}dg,$ $\iota_v[\alpha]\equiv df[v]{\wedge}dg{-}df{\wedge}
dg[v]=v^i(\pa_i f dg-df\pa_i g),$ etc. This generalizes in an obvious way to arbitrary 
p-forms. Like the exterior differential, this is a coordinate-independent operation 
having the property that $\iota_v^2\alpha=0$ on any form $\alpha.$

\section{The $SO(5)$ spherical harmonics}
\label{sec:sphh}

We very briefly review the $SO(5)$ spherical harmonics $(n,0),$ which are
all that are needed in this paper.  (For further discussion, see 
\reference{Gibbons:1978ji} and references therein.)  In Cartesian coordinates, these 
functions are formed from the five-vector $x^i:$  
$f_n(x) \equiv \linebreak
f_{i_1i_2\ldots i_n} x^{i_1i_2\ldots i_n},$ with $f_{i_1i_2\ldots i_n}$ 
a constant, symmetric, traceless co-tensor.  $f_0$ is just a constant. $f_1$ takes 
the form $f_1=f_k x^k.$ $f_2=f_{ij}x^ix^j,$ with $f_i{}^i=0,$ etc. These can be 
associated with the irreducible representations (irreps) frequently labeled by their 
dimensions: $\{\bf{1, 5, 14, \ldots }\}.$ These irreps are harmonics, i.e., solutions 
of Laplace's equation in five dimensions, $\Box_5 f_n=0.$ (For Lorentzian 
signature, the d'Alembertian replaces the Laplacian.)  

In spherical coordinates, these take the form 
$f_n(r,\omega^\alpha)=r^n\phi_n(\omega^\alpha),$ where $\omega^\alpha$ 
denotes the four angles implicitly defined in \eqn{eq:spherical}.  Using 
the metric from \eqn{eq:sphmetric}, we may write Laplace's equation as
\beq
-\Box_5 f_n=\left[-\frac{1}{r^4}\frac{\pa}{\pa r}\left(r^4\frac{\pa}{\pa r}\right) 
+\frac{1}{r^2} L^2\right]r^n \phi_n(\omega^\alpha) =0,
\eeq
where $L^2$ denotes the quadratic Casimir\footnote{$L^2$ is sometimes  called 
the spherical Laplacian or the Laplace-Beltrami operator on the sphere.} of 
``orbital" angular momentum in five-di\-men\-sions. Explicitly, 
$L^2\equiv\sum_{i<j}L_{ij}^2,$ with $L_{ij}dx^i{\wedge}dx^j$ the 10 generators of 
$SO(5).$ In Cartesian coordinates, $L_{ij}=-i(x_i\pa_j-x_j\pa_i).$  Carrying 
out the radial derivatives and evaluating on the $S^4$ submanifold $r=r_0$ yields
\beq
- \Box_{S^4} \phi_n^m(\omega^\alpha)=
\frac{1}{r_0^2}L^2  \phi^m_n(\omega^\alpha)=
\frac{n(n+3)}{r_0^2} \phi^m_n(\omega^\alpha),
\eeq
where $r_0^2=12/R_0,$ $R_0$ being the curvature of $S^4,$ and $m$ labels the 
linearly independent functions having a common eigenvalue. Thus, the 
``spherical harmonics" $\phi_n^m(\omega^\alpha)$ obey 
\beq
L^2\phi_n^m(\omega^\alpha)=n(n+3)\phi_n^m(\omega^\alpha).
\eeq
These symmetric irreps are sometimes called the $(n,0)$ representations 
(because a second integer $(n,p)$ is needed to delineate all representations).  
It is a combinatoric exercise~\cite{Hamermesh:grpth} to determine that the 
degree of degeneracy of eigenvalue $n(n+3)$ is 
\beq
d_n=\frac{1}{6}(n{+}1)(n{+}2)(2n{+}3),
\eeq
the dimension of the representation.  As a check, the non-isometric zero modes 
correspond to $n=1$, for which $d_1=5.$  

On $S^4,$ with $r=r_0,$ the Cartesian coordinates are not intrinsically well-defined, 
but we may continue making reference to the ambient space by using the angular 
variables $\vartheta_\alpha$ to label points on $S^4.$  In other words, since 
$x^ix^j\delta_{ij}=r_0^2,$ only four of the five coordinates $x^i$ are 
independent on $S^4.$  Similarly, we may continue using the vierbein 
$e_\alpha,$ \eqn{eq:funfbein}, as a local basis of the tangent space $TS_{\bf x}^4.$ 
Just as we denoted $e_r$ by $e_r^i=\hat{x}^i,$ it is convenient to continue using five-
component notation for $e_\alpha^i$ in order to avoid having to specify the choice of 
coordinates on $S^4.$  Further, the $e_r^i=\phi^i_1(\omega^k)$ do transform five-
vectors under $SO(5)$ rotations.

We have pointed out in \eqn{eq:hodge} that the Hodge dual of $e_r$ in $E^5$ 
is proportional to the four-form associated with the volume 
$dV_4$ on $T^*\!S^4.$ Since $dV_4$ is coordinate invariant, it is a gauge invariant, 
but the identification with the conformal rescaling of the metric in the unitary TT-
gauge is gauge dependent.  For example, in the unimodular gauge, $\sqrt{g_4}=1,$ 
the association would be quite different.  The conformal Killing equation however is 
gauge covariant.

\section{Killing vectors and Killing forms}
\label{sec:kill}

In this section, we elaborate on the disposition of the five Killing vectors on $E^5$ 
resulting from translation invariance when restricted to an $S^4$ submanifold.  
We know that there are five zero modes of the fluctuations on $S^4,$ but we wish 
to understand how they might be related, if at all, to the isometries of $E^5,$ the 
Poincar\'e semigroup.  Translation invariance is manifest in Cartesian coordinates,
and the corresponding Killing equations are 
\beq\label{eq:cartkill}
\nabla_i \xi_j+\nabla_j \xi_i=0.
\eeq
Since the metric is simply $\delta_{ij},$ the spin connection vanishes, so that 
$\nabla_i \xi_j=\pa_i \xi_j.$
This implies that the five components $\xi_i=k_i$ for arbitrary constants  
$k_i$ generate translations.  Alternatively, since the inverse metric is $\delta^{ij},$ 
we may say that $k^i\pa_i$ is a Killing vector for an arbitrary constants $k^i.$  
However, we are interested in $S^4,$ which is not translation invariant, and it is 
not at all clear whether any of these are projected onto Killing vectors on the 
$S^4$ submanifold. 

To facilitate the connection between $E^5$ and $S^4,$ let us rewrite the 
Killing equations, \eqn{eq:cartkill}, in spherical coordinates, \eqn{eq:spherical},
\begin{subequations}\label{eq:sphkill}
\begin{align}
\label{eq:sphkillrr}
\nabla_r \xi_r&=0,\\ 
\label{eq:sphkillra}
\nabla_r\xi_\beta+\nabla_\beta\xi_r&=0,\\ 
\label{eq:sphkillab}
\nabla_\alpha\xi_\beta+\nabla_\beta\xi_\alpha&=0.
\end{align}
\end{subequations}
(Recall that, with the exception of the radius $r$, we use Latin indices for 
Cartesian coordinates, and Greek indices for spherical coordinates on $S^4.$ 
In \eqn{eq:sphkill}, we have abbreviated the 
angular components $\omega^\alpha,$ with a slight abuse of notation, simply by 
the index $\alpha.$)  Although the submanifold of special interest has radius $r=r_0,$ 
for the time being, we can take any fixed value of $r.$  In 
spherical coordinates, the connection is non-trivial. Noting the metric 
\eqn{eq:sphmetric}, the nonzero connections in spherical coordinates take the form 
\beq\label{eq:sphconn}
\Gamma_\ab{}^r=-\frac{g_\ab}{r},\quad 
\Gamma_{r\beta}{}^\alpha=\frac{\delta_\beta^\alpha}{r},\quad 
\Gamma_\ab{}^\lambda,
\eeq
where $\Gamma_\ab{}^\lambda$ are the connections on the $S^4$ submanifold 
at fixed $r.$  Although their  precise form will not be needed, we note that 
$\Gamma_\ab{}^\lambda$ is  independent of $r$. 

The first equation above,
\eqn{eq:sphkillrr}, becomes $\pa_r\xi_r=0,$ so that $\xi_r=\xi_r(\omega^\alpha),$ 
independent of $r.$ This agrees with our orthonormal basis in $E^5,$ 
\eqn{eq:funfbein}, since $k_r=k_ie^i_r(\omega^\alpha)$ is independent of $r.$ 
Using this in the second equation, \eqn{eq:sphkillra}, we see 
that $\xi_\beta$ must be linear in $r,$ in agreement with 
$\xi_\beta= k_ie^i_\beta.$  Thus, $\nabla_r\xi_\beta=\pa_r\xi_\beta-\xi_\beta/r=0.$
Therefore, this equation implies that each covariant derivative vanishes separately, 
$\nabla_\beta\xi_r=\pa_\beta\xi_r-\xi_\beta/r=0,$ which agrees with \eqn{eq:funfbein}.
Finally, \eqn{eq:sphkillab} is not quite the same as the corresponding equations on 
$S^4,$ because the connection on $T^*\!E^5$ differs from the connection on 
$T^*\!S^4,$\footnote{\eqn{eq:gauss} is a special case of 
Gauss's equation in which the second term on the right-hand side is associated 
with the second fundamental form on $S^4$ embedded in $E^5.$}
\beq\label{eq:gauss}
\nabla_\alpha\xi_\beta=\left(\nabla_\alpha\xi_\beta\right)_{\!4}-\Gamma_\ab{}^r\xi_r
=\left(\nabla_\alpha\xi_\beta\right)_{\!4}+\frac{g_\ab}{r}, 
\eeq 
where, by definition, $\left(\nabla_\alpha\xi_\beta\right)_{\!4}$ only involves the $S^4$ 
connection $\Gamma_\ab{}^\lambda.$  Thus, \eqn{eq:sphkillab} becomes 
\beq\label{eq:fourkill}
\left(\nabla_\alpha\xi_\beta+\nabla_\beta\xi_\alpha\right)_{\!4}=-2\frac{g_\ab}{r}.
\eeq
This  implies $\left(\nabla_\alpha\xi^\alpha\right)_{\!4}=-4/r,$ so that the preceding 
equation may also be expressed as 
\beq
\left(\nabla_\alpha\xi_\beta+\nabla_\beta\xi_\alpha\right)_{\!4}=
\left(\nabla{\cdot}\xi\right)_{\!4}\frac{g_\ab}{2}.
\eeq
We may set $r=r_0$ here\footnote{These relations hold for \emph{any} fixed value 
of $r,$ an observation that proves useful in classically scale-invariant 
models~\cite{Einhorn:2014gfa, Einhorn:2015lzy, Einhorn:2016mws}, wherein the 
value of the scalar curvature is classically undetermined.} to conclude that the 
Killing forms for infinitesimal translations $\xi_\alpha$ on $T^*\!E^5$ project onto 
non-isometric, conformal Killing forms on $T^*\!S^4.$ 

The components $\xi_\alpha$ of a co-vector in spherical coordinates 
are related to the corresponding co-vector $k_i$ in Cartesian coordinates 
according to $\xi_\alpha=k_i e^i_\alpha.$  Since $k_i$ are arbitrary constants, 
this implies that there are five conformal Killing forms 
$e^m_{\vartheta\!_\beta} d\vartheta_\beta$ on $T^*\!S^4,$ where the index $m$ 
may be identified with the Cartesian components~$e^m_\beta.$

The preceding concerns the properties of the dS background and not the 
fluctuations directly, but the generators of isometries have consequences for  the 
infinitesimal fluctuations and implications for zero modes. We have 
seen that none of the translation generators $\xi^\alpha$ in the tangent bundle 
$TS^4$ are true Killing vectors, so they do not generate isometries.  (The 
interpretation of $\xi_r$ for fluctuations on $S^4$ is discussed in \secn{sec:ds5}.)  
Therefore, none of these observations imply that any fluctuation is directly 
associated with a zero mode of the action on $S^4,$ but we have presented such 
an argument in \secn{sec:5zero}. 

That argument did not require this result on the conformal Killing forms although it 
is a corollary. Even though these $e^m_{\vartheta_j} d\vartheta_j$ do correspond to 
zero modes, they are not really new. Recall that the most general Killing form on 
$E^5$ is $\omega_{ij}x^j+k^i,$ with the antisymmetric constants $\omega_{ij}$ 
corresponding to the 10 rotation generators of $SO(5)$.  
For a fixed direction $x^j,$ there appear to be five nontrivial rotation generators 
$\omega_{ij}x^j,$ but, since the radial projection onto $e^i_r$ vanishes, there are 
only four non-trivial rotations\footnote{The little group of a fixed point on $S^4$ 
is $SO(4),$ the 6 generators that annihilate the normal $\hat{x}^i=e^i_r$ at that 
point.  This leaves $10-6=4$ non-trivial rotations isomorphic to the cosets  
$SO(5)/SO(4).$} at fixed $x^i$ on $S^4.$  We may denote them by 
$\omega_{r \vartheta_k}=\omega_{ij}e^i_re^j_{\vartheta_k}=-\omega_{\vartheta_k r}.$ 
At fixed $r,$ the rotation group $SO(5)$ remains a good symmetry, so these four 
rotations do reflect true Killing forms at fixed $x^i.$ and, therefore, do 
correspond to zero modes. Even though the translations $\xi_i$ are not  
isometries on a fixed $r$ submanifold, their projection onto the four-sphere of 
radius $r$ can be compensated by a rotation, viz., one may choose 
$\omega_{r \vartheta_k}$ such that
\beq\label{eq:comp}
\left(\xi_j+\omega_{ij}r e^i_r\right)e^j_{\vartheta_k}=
\xi_{\vartheta_k}+r\omega_{r \vartheta_k}=0,
\eeq
for each $\vartheta_k.$  Paradoxically, the conformal Killing forms $\xi_\alpha$ may 
locally be written as a sum of true Killing forms.  Since the linear combination 
depends upon the direction $\hat{x},$ this is understandable. The really surprising 
result is encoded in the five ``radial" zero modes which, we argued, reflect 
displacement of the center of the $S^4$ sphere.

\vfill

\pagebreak

\end{document}